# An Online Deep Learning Approach Toward the Prediction of Power System Stresses Using Voltage Phasors

Elham Foruzan, *Member, IEEE*, Sajjad Abedi, *Member, IEEE*, Jeremy Lin, *Senior Member, IEEE*,
Sohrab Asgarpoor, *Senior Member, IEEE,* and Emanuel Bernabeu

*Abstract*— The outage of a transmission line may change the system phase angle differences to the point that the system experience stress conditions. Hence, the angle differences for post-contingency condition of a transmission lines should be predicted in real time operation. However, online line-based phase angle difference monitoring and prediction for power system stress assessment is not a universal operating practice yet. Thus, in this paper, an online power system stress assessment framework is proposed by developing a convolutional neural network (CNN) module trained through Deep Learning approach. In the proposed framework, the continuously streaming system phase angle data, driven from phasor measurement units (PMUs) or a state estimator (SE), is used to construct power system stress indices adaptive to the structure parameters of the CNN module. Using this approach, any hidden patterns between phase angles of buses and system stress conditions are revealed at low computation cost while yielding accurate stress status and the severity of the stress. The effectiveness and scalability of the proposed method has been verified on the IEEE 118-bus and more importantly, on the PJM Interconnection system. Moreover, outperformance of the proposed method is verified by comparing the results with artificial neural network (ANN) and decision tree (DT).

## I. Introduction

The importance of monitoring phase angle differences to determine the system stress state for the pre- and post-contingency of power systems has been highlighted in reports analyzing the 2003 North American blackout and the 2011 Pacific Southwest outage [1-2]. The report of the 2011 Pacific Southwest outage states that following the forced outage on a 500 KV line, the phase angle difference of two terminal buses increased drastically, which eventually led to cascading outages and forced 2.7 million customers out of power [3]. Additionally, the 2003 North American blackout report uncovers the lack of phase angle awareness as an important reason for the blackout. Based on these reports, the North American Electric Reliability Corporation (NERC) emphasizes on developing tools for *monitoring* angle difference in real-time, as they provide the system operator with an immediate awareness of the system stress. System stress simply refers to the amount of overloading that transmission lines carry; this concept will be explained in more details in this paper [3-4]. However, the literature on methods toward exact online *prediction* of power system stress is quite immature.

Currently, system operators use commercial real-time contingency analysis (RTCA) tool to obtain awareness of the phase angle differences and thermal line overloads. The RTCA commonly uses state estimation (SE) data as the input and runs AC power flow equations every five minutes to calculate the post contingency snapshot of these variables for all credible contingencies in the current Operating Condition (OC) [5-7]. RTCA calculation is not fast enough and suitable for online applications, especially for large-scale and highly loaded systems. This leaves the operators with no choice but to rely on approximation methods such as using the process information (PI) displays to monitor system situation. However, PI displays are built based on pre-calculated distributed factors (DFs) in the planning horizon for offline studies [7]. Relying on DF values decreases the accuracy and produces unreliable results, as real-time system topology and OCs may be different from those in the planning horizon. Moreover, both RTCA and PI display uses SE data to monitor system, thus in the case of SE failure, operators do not have any alternative tool to monitor system post-contingency status.

Along this venue, the usage of emerging phasor measurement unit (PMU) technology is going more prominent to improve the capability of grid operations toward fast online monitoring over a wide-area footprint of bulk power systems [8]. The PMU time-synchronized data can be used for assessing system stability following a major event, and alerting system operators to access precise real-time results within seconds of a system event. Given this opportunity, it would be helpful to have a tool that can transform the PMU raw phase angle dataset into meaningful information presentable to system operators. This new tool can also be used in the case of SE failures. Some advances in the use of PMU data to monitor system stability and stress are described in [9-10]. Dobson in [9] used a model reduction method based on "cutset area selection" to reduce system dimensions. This method used the phase angle of a limited number of buses to monitor system stability in real-time operation. Wu et al. in [10] selected the critical generators buses to evaluate the transient stability of a power system. However, these approximations to limit the number of monitored buses may decrease accuracy and produce less reliable results. Hence,

Elham Foruzan is with the Electrical and computer Engineering Department at University of Nebraska, Lincoln, NE, USA and California Independent System Operator, Folsom, CA, USA (elham.foruzan@huskers.unl.edu).

there is a pressing need to develop fast and reliable online modules for identifying OCs leading to stressed states in a highly uncertain operating environment, where accuracy is not sacrificed in favor of computational requirements. The objective in this paper is to develop a tool that processes a large amount of phase angle data driven from *all PMUs* and provide more meaningful and concise information about system stress and strength states.

To address the challenge of time complexity and achieve faster system assessment for online applications, *empirical solutions* based on machine learning (ML), replacing *analytical/numerical solutions*, have been adopted that have a high scalability and an ability to map complex patterns based on available ceaseless data measurements. A common capability of this approach is that it can take the measurements of state variables as input and directly map the defined security metrics in the output, circumventing the cumbersome numerical models. In this area of research, artificial neural networks (ANN) [11-14], decision trees (DT) [15-19], and support vector machines (SVMs) [20] are employed as tools for system security assessment. For instance, ANN-based methods for online ranking of contingencies based on their severity level are presented [11, 12]. Other researchers presented DT-based methods with traceable multivariable output that gives more information to the operator, as opposed to a black box model of an ANN with a uni-variable output [15-16]. For example, a regression tree approach was studied in [15] for predicting the power system stability margin and predicting the ex-post events. These studies mostly considered monitoring pre-contingency states as an indicator of near-future system security conditions. However, the line-based phase angle difference monitoring, which is a new power system concern after the 2003 North American blackout and 2011 Pacific Southwest outage [1-2], has not been addressed. Another drawback of such ML approaches is the need to perform 'data preprocessing stage in order to select appropriate training attributes from empirical data, which loses the model generality and requires the design of the model specific to the system under study.

In this paper, an online power system stress prediction framework based on a Deep Learning (DL) convolutional neural network (CNN) module is proposed that is capable of monitoring and processing a large amount of phase angle difference data obtained from PMUs and provides precise prediction on the status and severity of system stress conditions. In summary, the contributions made in this work are listed as follows:

- The developed framework based on the DL CNN overcomes the large computational overhead of real-time phase angle monitoring while achieving high accuracy results. CNN is reported to outperform other ML methods in automatic feature extraction from complex and massive data [23-27]. The CNN has produced remarkable results in pattern recognition for a variety of problems, including automatic image style recognition [23-24], natural language processing [25], and classification of hyperspectral data [26]. A self-improving CNN is reported to have successfully solved the *'curse of dimensionality'* by iteratively selecting the most informative input data [27]. In these applications, various forms of CNN have been reported to improve the accuracy of predictions while significantly reducing the computational time of the existing solutions.
- In this study, simple formulations are tuned to identify the extent to which the imminent OC is vulnerable to the contingencies using the phase angles of buses fed by PMUs or state estimator (SE)/)/supervisory control and data acquisition (SCADA). Using this input data, the CNN is to map the output in the form of system stress status and severity for the current OC. Therefore, it is noteworthy that the proposed method can be used as an alternative for RTCA in the case of failure in SE.
- Other than case studies on the IEEE 118-bus test system, the performance and scalability of the proposed method is demonstrated in practical applications through exhaustive test on the PJM Interconnection system. The results of our case studies testify that the proposed framework can accurately predict the near-future system stressed states based on the phase angle in the PJM network, while significantly reducing the computational time. Further, comparative case studies using models based on conventional ML methods, namely ANN and DT, are conducted on the PJM Interconnection power system indicating outperformance of the proposed method in terms of accuracy and the ability to predict any possible stress conditions.

## II. PHASE ANGLE DIFFERENCES AND SYSTEM STRESS CONDITIONS

The measured phase angle difference between buses in bulk power systems is an indication of system stress condition [3]. Therefore, it can be used to identify early signs of system stress after sudden changes in system topology or power flow. In this section, we first describe the link between system phase angles and the line power transfer. Then, a systematic criterion are presented to assess the system stress.

### A. Fundamental Drivers for Phase Angle Difference

The phase angles of buses are strongly related to active power transfer and system topology. The relationship between system phase angles and power transfer in a transmission line is represented by the power transfer equation [28]:

$$P_{rs} = \frac{V_r V_s}{X_{rs}} \sin(\theta_{rs}) \tag{1}$$

where $V_r$ and $V_s$ are the voltage magnitude of the receiving and sending end buses, respectively. $X_{rs}$ is the transmission line impedance and $\theta_{rs}$ is the phase angle difference between the terminal buses of the line. Usually, in bulk power systems, voltage magnitude is held relatively close to 1.0 pu within a reasonable operating tolerance. Hence, phase angle difference is directly related to power flow and electrical impedance. According to Eq. (1), a large phase angle difference between a pair of buses indicates greater power flow between those points, which may result to higher stress across the transmission line.

When major events in a power system, such as transmission and generation outages or contingencies take place, the system topology may dramatically change, which in turn alters the equivalent $X_{rs}$ between any bus pair ($r, s$). These events may increase the bus phase angle differences, especially for buses electrically close enough to the events. Thus, phase angle differences are not only indicators of power flows, but also a strong indicator of system topology conditions and switching events. The relationship described here is a fundamental link of phase angle differences of system buses and transmission line stress, which is used in the proposed online stress assessment framework in this paper.

*B. Online Static Stress Assessment*

The purpose of the proposed system stress assessment framework in this paper is to provide sufficient information to the operators on the status of line loading and system stress states in the event of any contingency, based on the current OC's phase angles.

For each line $i \in M$-lines, $PF_i^{SL}$ is defined to represent a stress limit (SL) for line $i$. Based on this limit, the operating stress for a given transmission line $i$ is defined as:

$$TL_{i,s} = \begin{cases} [|PF_i| - PF_i^{SL}]/PF_i^N & |PF_i| > PF_i^{SL} \\ 0 & otherwise \end{cases} \quad (2)$$

where, the $PF_i$ and $PF_i^N$ are the active power and normal rating for line $i$. Based on this definition, a given power system is considered to be under stress if the loading of any line $i \in M$ exceeds its designated stress limits. Then, an alarm limit (AL), $PF_i^{AL}$, for line $i$ adjacent to the stress limit is assigned to give more information and flexibility for performing preventive [29].

In this paper, the Security Index (*SI*) limit in terms of a ratio based on the line flow is used for online system stress assessment [29]. Considering each line's stress and alarm limits, system stress index (*SI*) is described by Eq. (3) to distinguish the system unstressed, alarm and stressed conditions. The *SI* for *M*-lines in the system is defined as:

$$SI = \left[\sum_{i=1}^{M}\left(\frac{d_i}{g_i}\right)^{2n}\right]^{1/2n} \quad (3)$$

Where $g_i$ is normalizer and $d_i$ is the normalized power deviation from the alarm limit:

$$\begin{cases} d_i = [|PF_i| - PF_i^{AL}]/PF_i^N & |PF_i| > PF_i^{AL} \\ d_i = 0 & otherwise \end{cases} \quad (4)$$

The normalizer $g_i$ is defined as in Eq. (5):

$$g_i = \frac{[PF_i^{SL} - PF_i^{AL}]}{PF_i^p} \quad (5)$$

Here we define system stress state as a binary variable indicating if the post-contingency of OC is stressed or unstressed. If there is no line with a flow greater than its stress limit, then *SI* will be *zero* and system stress state is zero; and the system is designated in unstressed condition. In contrast, if in the post-contingency condition, any line is loaded over its stress limit, the index value will be greater that one; and the system is considered in stressed condition. In this case, system stress state is equal to one. Values between zero and one represent alarm conditions.

## III. ONLINE SYSTEM STRESSED ASSESSMENT MODULE BASED ON CNN

To obtain fast and accurate system stress status, an online intelligent pattern recognition module is developed in this paper, capable of capturing the hidden patterns between phase angles of buses and system stress conditions, as defined in the previous section. The CNN module is developed in two stages as depicted in Fig. 1. In the first stage, a database with sufficient OCs, from either off-line simulation or selected from the PMU dataset, is created. The OCs in the database should be capable to represent the entire range of OCs of the system under study. In the second stage, the CNN is trained using the database created in the first stage. The developed CNN module is trained to characterize the stress status or predicted SI for any OC by mapping from the phase angle of the buses. Fig. 2 shows the structure of the proposed online system stress assessment module. The processes for generating OCs and the details of implementing a CNN module are explained in the remainder of this section.

*A. Pattern Generation*

To create a reliable database, a complete set of $N_{oc}$ OCs by off-line full Newton Raphson (NR) power flow simulation at different times of the day corresponding to various load profiles is created. In the offline approach, from the *N* buses in the system, the $N^2$ pairs of voltage angle differences are collected. Then, all the $N_c$ contingencies, here we consider all of the N-1 contingencies, including all generators and transmission lines, are applied to each OC one at a time; and the post-contingency power flow is solved. Finally, the SI is calculated based on Eq. (3) for each OC and contingency pair.

After applying a set of $N_c$ contingencies to all the $N_{oc}$ OCs, a database of $N_c \times N_{oc}$ patterns is generated. For each pattern, a vector consisting of $N^2$ voltage angle differences and an applied contingency $I_d$ is formed as an input attribute vector fed to the CNN. The output of CNN can be the real value of calculated *SI* or binary value that represent system stress status.

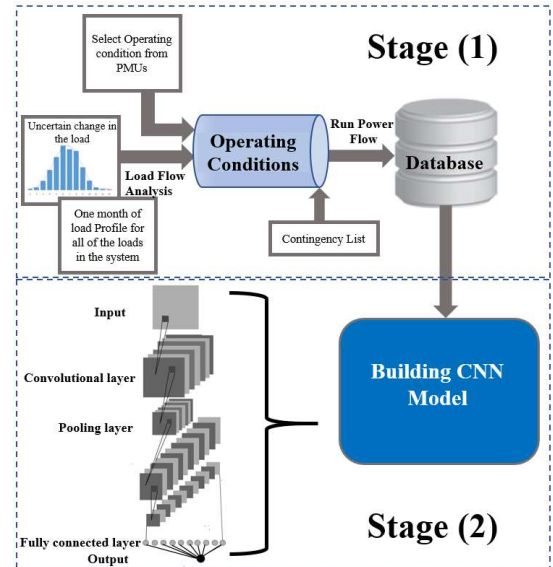

Figure 1. Flowchart of the proposed online power system stress assessment (OSSA) framework.

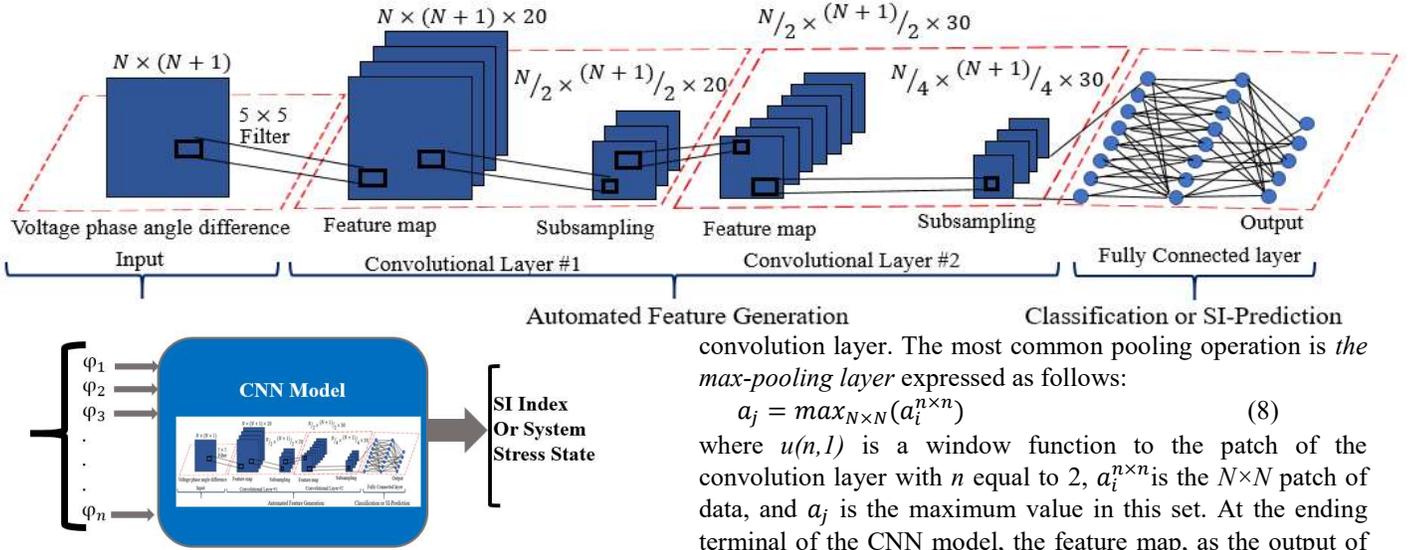

Figure 2. Structure of the deep learning CNN at stage 2.

## B. The Deep Learning CNN Module

The major focus of the proposed CNN module is to recognize complex patterns from the empirical data and map the data to generate outputs which intelligent decisions can be made upon. We illustrate how the proposed intelligent module using CNN maps the operating point phase angles of buses into the prediction model of stresses in this section.

The architecture of CNN is a combination of consecutive layers, each applying a nonlinear complex function to the data fed from the previous layer. Fig. 3 illustrates a typical feed-forward CNN network. In general, each layer applies a nonlinear transformation to its input and provides automated feature selection to its output to the next layer, consecutively. The CNN consists of one or more convolutional layers which convolves its input with a number of local filters or kernels to extract different features of the input and form multiple feature maps for feature mapping (or feature representations) [30-31]. Following each convolutional layer is an element-wise nonlinearity, which allows the CNN to learn other relevant nonlinear mappings [30-31]. This process is described by Eq. (6). The value of a neuron $v_{i,j}^x$ at position $x$ of the $j^{th}$ feature map in the $i^{th}$ layer is calculated as:

$$v_{i,j}^x = g\left(b_{i,j} + \sum_m \sum_{p=0}^{P_i-1} \omega_{i,j,m}^x v_{(i-1),m}^{x+p}\right) \quad (6)$$

where $m$ is the feature map in the $(i-1)^{th}$ layer; $\omega_{i,j,m}^x$ is the weight of position $p$ connected to the $m^{th}$ feature map; $P_i$ is the width of the kernel in the spectral dimension; and $b_{ij}$ is the bias of the $j^{th}$ feature map in the $i^{th}$ layer. The function $g(.)$ is a nonlinear function which introduces nonlinearity into the model. The rectified linear units (*ReLU*) activation function is the proper choice as the nonlinear function:

$$g(x) \doteq Relu(x) = max(0, x) \quad (7)$$

A pooling layer is typically inserted in between successive convolutional layers to reduce the spatial size of the parameters in the network and for controlling overfitting. Each pooling layer corresponds to its preceding convolutional layer. The neurons in the pooling layer combine a $N \times N$ patch of the convolution layer. The most common pooling operation is *the max-pooling layer* expressed as follows:

$$a_j = max_{N \times N}(a_i^{n \times n}) \quad (8)$$

where $u(n,l)$ is a window function to the patch of the convolution layer with $n$ equal to 2, $a_i^{n \times n}$ is the $N \times N$ patch of data, and $a_j$ is the maximum value in this set. At the ending terminal of the CNN model, the feature map, as the output of the last max-pooling layer, is fed into the *fully-connected* layer where the neurons are all connected to all activation units in the previous layer. The *fully-connected* layer is used for prediction during the training process, as it is capable of combining the features abstracted from previous layers for final binary classification or real value *SI* prediction. Finally, to reduce overfitting, *"dropout"* is applied in the fully connected layer. In this step, the last layer samples out different architectures by using the probability distribution function used in the fully connected layer during training process. Therefore, some neurons that are present in the *fully-connected* layer may be removed in the next training iteration.

Weights among all layers, including the convolutional layers and the fully connected layer of the CNN model, are trained using a backpropagation algorithm and an adaptive moment (Adam) optimization algorithm [22]. The backpropagation algorithm is an iterative algorithm to determine the parameters of CNN model that minimize the cost function.

Adam [22] is a method that computes adaptive learning rates for each parameter in each iteration $t$ of backpropagation algorithm. Adam stores both the exponentially decaying average of past gradients $m_t$ and past squared gradients $v_t$. $m_t$ and $v_t$ are estimates of the first moment (the mean) and the second moment of the gradients respectively:

$$m_t = \beta_1 m_{t-1} + (1 - \beta_1) g_t \quad (9)$$
$$v_t = \beta_2 v_{t-1} + (1 - \beta_2) g_t^2 \quad (10)$$

where $g_t$ is gradient of cost function at time $t$.

The cross-entropy (*Cr*) and Mean Squared Error (*MSE*) Loss are used as a cost function for binary classification and real value prediction as represented by Eqs. (11) and (12):

$$Cr = -\frac{1}{n}\sum_{i=1}^{n}[Y_i \ln Y_i' + (1-Y_i)\ln(1-Y_i')] \quad (11)$$
$$MSE = \frac{1}{n}\sum_{i=1}^{n}(Y_i - Y_i')^2 \quad (12)$$

Where $n$ is the total number of items of training data, $Y_i'$ and $Y_i$ are the predicted and desired output corresponding to input $i$. Finally, all parameters update in iteration $t+1$ as follows:

$$\theta_{t+1} = \theta_t - \eta \frac{m_t'}{\varepsilon + \sqrt{v_t'}} \quad (13)$$

where $m'_t$ and $v'_t$ are formulated in Eqs. (15) and (16):

$$m'_t = \frac{m_t}{1-\beta_1} \quad (14)$$

$$v'_t = \frac{v_t}{1-\beta_2} \quad (15)$$

The values of $\beta_1$ and $\beta_2$ are set to 0.9 and 0.999, respectively, and $\varepsilon$ is set to a small value $\sim 10^{-9}$ [22].

Finally, the accuracy of CNN model calculated using mean absolute percentage error (MAPE) as in (16) and (17):

$$\text{MAPE} = \frac{\sum_i |Y_i - Y'_i|/Y_i}{n} \quad (16)$$

$$Accuracy = 1 - MAPE \quad (17)$$

Once the CNN module is trained, it will be able to take a particular OC in terms of the phase angles of all buses as the input, and provide the system stress state or the *SI* value as the output, as defined in Eq. (3). In this study, the CNN is configured consisting of two convolutional layers to select the best set of features from the input data and two *fully-connected* layers to perform the final classification. In the first convolution layer, ten 5 x 5 filters (extracting 5 x 5 data set) are applied with the *ReLU* activation function to extract local features of bus phase angles. Then, within the same layer, max pooling layer with a 2 x 2 filter and a stride of 2 are applied. In the second convolutional layer, twenty 5 x 5 filters are applied, with the *ReLU* activation function. Again, *max-pooling* layer with a 2 x 2 filter and a stride of 2 is used. The *fully-connected* layer consists of 1,100 neurons, with a dropout regularization rate of 0.5, which corresponds to the probability of 0.5 that any given element will be dropped out during training. Finally, the output layer has a node to calculate the stress state or SI as in Eq. (3).

## IV. NUMERICAL RESULTS

The effectiveness of the proposed method is verified on the IEEE118-bus system and results are analyzed to confirm the accuracy and computational speed of the proposed online system stress assessment framework. Then, the framework is applied to the PJM Interconnection system and compared with two widely used ML models, namely ANN and DT.

### A. IEEE 118-Bus Test System

The IEEE 118-bus system consists of 54 generators, 177 transmission lines, and 9 transformers [32]. A total number of $N_c = 240$ contingencies including lines, transformers, and generator outages are considered and simulated in the case study. The alarm and stress limits for line flows are considered to be 90% and 110% of nominal rating. Here, we created a total of $N_{oc} = 14400$ OCs for 30 days, where load data in the system is updated every three minutes [33]. In this way, 480 system snapshots is created within one day, leading to 480 different system OCs. The same process is repeated for 30 days. In addition, to consider the uncertainty on the OCs in this process, a Monte Carlo simulation of normal distribution is generated, and values are added to each load for every OC.

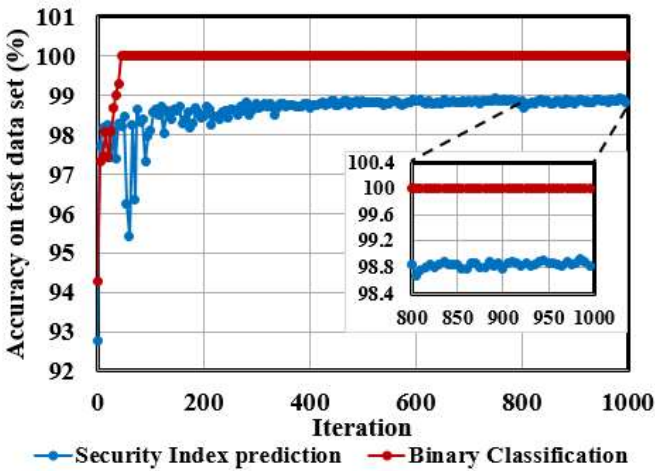

Figure 4. Accuracy of SI and stress status results on the test dataset versus the number of training epochs in the IEEE 118-bus case study.

TABLE I ACCURACY RESULTS OF CNN ON FIVE-FOLD CROSS VALIDATION

| $K^{th}$ Fold | 1 | 2 | 3 | 4 | 5 |
|---|---|---|---|---|---|
| Predictive CNN Accuracy (%) | 98.80 | 98.92 | 98.76 | 98.64 | 98.86 |
| Classification CNN Accuracy (%) | 100 | 100 | 100 | 100 | 100 |

The training dataset includes 70% randomly selected points of the produced dataset, and the remaining 30% is used as the test dataset to evaluate the model. The accuracy of the model is evaluated with the test dataset in 1,000 iterations as depicted in Fig. 2. In each training iteration, the model is rigorously tested with the test dataset. As observed in Fig. 4, the classification CNN layer reached an accuracy of 100% on the test dataset in less than 60 iterations. This result approves that the CNN is structured such that it can learn very fast and achieve a high accuracy of 100% on a test dataset. Thus, it is able to correctly classify all of stressed and unstressed operation states, which provides valuable and completely reliable outcomes to the utility operators.

Additionally, Fig. 4 shows that the performance of the CNN trained to predict the real value *SI*s increases with the increasing training iterations but converged to a high accuracy of 98.8% on the test dataset at the 500[th] iteration. This result demonstrate that the CNN module is quite capable of predicting the status and severity of post-contingency stresses with a very high accuracy. The required training time for the stressed state classification and *SI* prediction models is measured to be 56 and 435.66 seconds, respectively. In real time application, the CNN module retrains periodically, every hour to capture new changes, e.g. building new transmission line or transformer and train with all historical OCs.

To further ensure the robustness of the proposed method in delivering the same accuracy for various dataset scenarios, the trained CNN model is tested on four additional test datasets of the database using *K*-fold cross-validation. Here, *K* is set to 5; and the results are presented in Table I. The test results for various datasets reveal that the average accuracy of the CNN model for classification and SI prediction for these datasets is 100% and 98.79 %, respectively. Hence, t the high quality of the results is a strong confirmation of the effectiveness of the proposed method.

The outstanding high accuracy results obtained from the proposed method can be explained in relation to the CNN's superior structure for exploiting spatial correlation using convolutional layers. This correlation mapping process is perfectly in accordance with the higher impact of topology change in electrically close areas as opposed to areas that are electrically far from the location of the component in contingency. In other words, the phase angle difference of any two buses changes nonlinearly with the event of a transmission line contingency, as shown in Eq. (1); and it is based on the Thevenin equivalent impedance between the two terminal buses in post-contingency system topology. One additional factor contributing to the high accuracy of the proposed framework is the capability of the CNN in automatic features extraction from complex and massive data. On this basis, the CNN makes the framework independent of human knowledge while providing complex representation (feature selection) of data patterns, which is the goal of accurate classification. Since these regional correlations change for different power systems with different number of buses and branch topology, it is very important that the convolutional layers in the CNN model can capture these local relationships adaptively in different systems and pass an appropriate dataset to the *fully-connected* layer. This characteristic also reduces the likelihood of overfitting in prediction and increases accuracy. As a result, the CNN configured in this paper is highly accurate in predicting power system stress conditions, which is an important feature for online system stress assessment.

TABLE II COMPUTATIONAL TIME TO PERFORM ONLINE SYSTEM STRESS ASSESSMENT ON THE IEEE 118-BUS SYSTEM

| Security Assessment Method | CNN Module | Traditional CA |
|---|---|---|
| Required Time (s) | 0.53 | 11.06 |

Computational time is another crucial requirement for online stress assessment in power systems. The total time required to train CNN is directly related to the size of the input dataset. In the 118-bus system case study, the CNN training time was 7.25 minutes. Table II shows the computation time required to conduct stress assessment using the proposed CNN module compared to conducting traditional contingency analysis (CA), when the system is loaded at 120% of the original load profile. The computational time for performing system stress assessment using the CNN module is significantly lower than the traditional method. The reason is attributable to the extra time that the traditional method needs to perform all of the credible contingencies, calculate the indexes for each contingency, and find the final stress condition. This process has a computational time of $O(n \log n)$, where $n$ is the number of buses for the system under study. In contrast, the CNN module is used to predict indexes which *'automates'* this process. Hence, the CNN module with its high accuracy and low computational time is well suited for online system stress assessment.

### B. PJM Interconnection System

To further evaluate the performance of the proposed method in practice, case studies on the PJM Interconnection power system, as a large-scale and pragmatic testbed, are conducted. Furthermore, results are compared with those from the models based on conventional ML methods, namely ANN and DT. This system has more than 15000 buses and holds nearly 400 PMUs at over 125 substations. In this testbed, we use the PMU data to train the CNN, ANN, and DT models. In training the DT model, the large input space of 400 × 400 causes overfitting. Thus, in order to have an identical input dataset for the three methods, which is handle able for DT, a sensitivity analysis is performed to select the most impactful substations on the system phase angles. Therefore, eight most observable substations from 125 substations at different locations of the system were selected. Then, the buses with higher voltage level are selected from PMUs with the same geographic locations.

*1) CNN Module*

In this study, a database with 38400 OCs was created through querying phase angles from the selected PMUs for 30 days of operation in the summer of 2016. Then, a contingency list of Nc=7890 is applied to each OC; and a total input dataset size of 38400 × 7890 is used to train and test the three models. The PJM Interconnection system is pretty stable given all credible contingencies, therefore, the alarm and stress limits for line power flows were set to 95% and 97% of nominal flow rating to have both unstressed and stressed cases in the database. In the following section, the results of the three models are discussed. Similar to the previous case study, two convolutional layers are used to further select the best set of features from the input data. The first and second convolution layers have five and nine filters, respectively. Both convolution layers have a stride of one and they are zero padded. A *max-pooling* layer over 2 × 1 blocks is used. The first *fully-connected* layer contains 1000 nodes and the output layer has one node to predict the system stress state.

Once the model is trained, the accuracy of the model is tested using *K*-fold cross validation, with *K*=20. As seen in Table III, the average accuracy of the CNN model is 100% based on the 20-fold validation. The results indicate that the security monitoring for the PJM Interconnection test system could be estimated with perfect accuracy by using the proposed method.

*2) Artificial Neural Network (ANN)*

The *ANN* consists of an input layer of size $\binom{n=8}{2} = 28$ nodes, two hidden layers of size 20 and 12 nodes, and an output layer with one output node [34]. The number of hidden layers and the nodes in each layer is selected based on the result of sensitivity analysis. The backpropagation algorithm with the Adam optimizer is used to update weights in every training iteration, which minimizes the error in the training stage. Table III shows the results of 20-fold cross validation for the ANN model. The average accuracy of the ANN model was 92.9%.

*3) Decision Tree (DT)*

The CART algorithm is used in this study to build a model based on DT classifier [35]. To avoid overfitting in the DT model, we performed pre-pruning to find the optimal value for the minimum number of splitting for each node (MNSN) which gains the maximum output accuracy for the test dataset. The sensitivity analysis is performed with MNSN ranging from 1 to 20. For each value of MNSN between 1 and 20, the *K*-fold cross-validation is applied, and the accuracy of the tree is calculated. The maximum accuracy equal to 86.8 % obtained when the MNSN equals to 12 samples. The results of *K*-fold cross validation accuracy, with *K*= 20, on a prune three is presented in Table III.

*C. Results Comparison for three models*

We perform a statistical Student's t-test to compare the three models, which were built with the same training dataset. The result of the Student's t-test showed that the CNN model had smaller error compared to the other two models, i.e. ANN and DT, with a confidence of nearly 100%. The CNN could find the hidden patterns inside the training data along with extracting features using convolutional layers, which eventually led to a more accurate model and avoided overfitting compared to the other two models. These layers automatically discovered the pattern between phase angles of buses that were electrically close to each other and passed the appropriate dataset to the fully connected layer. On the other hand, the ANN and DT structures, which searched for a pattern by generally looking at all of the phase angle differences in the system, were more prone to overfitting.

TABLE III ACCURACY RESULTS OF CNN, ANN AND DT IN THE PJM INTERCONNECTION SYSTEM

| $K^{th}$ Fold | CNN Test Accuracy (%) | ANN Test Accuracy (%) | DT Test Accuracy (%) |
|---|---|---|---|
| 1 | 100 | 86.3 | 79.6 |
| 2 | 100 | 72.7 | 42.7 |
| 3 | 100 | 96 | 85.9 |
| 4 | 100 | 100 | 100 |
| 5 | 100 | 99.06 | 87.4 |
| 6 | 100 | 100 | 99.4 |
| 7 | 100 | 85.4 | 84.3 |
| 8 | 100 | 42.18 | 21.3 |
| 9 | 100 | 100 | 100 |
| 10 | 100 | 100 | 100 |
| 11 | 100 | 98.75 | 100 |
| 12 | 100 | 100 | 100 |
| 13 | 100 | 99.16 | 87.8 |
| 14 | 100 | 100 | 96.5 |
| 15 | 100 | 84.24 | 94.7 |
| 16 | 100 | 95.1 | 94.7 |
| 17 | 100 | 100 | 100 |
| 18 | 100 | 98.58 | 89.58 |
| 19 | 100 | 100 | 100 |
| 20 | 100 | 100 | 100 |

*D. Confusion Matrix*

The training and testing datasets were not a balanced dataset, and the percentage of stressed states was significantly lower than unstressed states. This is because the cost of the system stress condition due to contingencies in real-time is very high, and normally avoided by the system operators. Usually, system operators consider post-contingency remedial actions, such as energy backup or switching, to maintain the system in a secure pre-contingency state. Therefore, in addition to accuracy as a metric of the real performance of the models, a confusion matrix-based interpretation of the results is conducted as a better representation when the dataset is unbalanced.

Tables IV, V, and VI show the confusion matrix for the three models that were trained and tested on an identical training and test data sets. The stressed and unstressed states correspond to positive and negative labels, respectively. As can be seen in Tables IV to VI, all three models had a high percentage of true negative (TN), when the system is unstressed; and the model predicted the unstressed state.

However, the high percentage of TN, which contributed to high accuracy in this problem, does not fully indicate the accuracy of the predictive model. As an instance, the selected test dataset had 14% stressed and 84% unstressed states. Thus, a model that merely predicts unstressed states for every input OC is 84% accurate. On the other hand, it is required for the model to identify potential stress conditions in advance and give system operators enough time to perform the preventive actions, as the cost of preventive actions is significantly lower than the cost of losing system stability. Therefore, it is desirable to have a model with a high true positive (TP) when the system is stressed, and the model predicts a stressed state; and low false negative (FN) when the system is stressed but the model predicts an unstressed state. The rate of FN in the ANN and DT predictive models is 46% and 40%, respectively, while this percentage of FN in the CNN model is *zero*. Thus, we can conclude that although the ANN and DT had an acceptable accuracy on the test dataset, they performed poorly on predicting situations leading to the stressed condition. On the other hand, the CNN model was capable to perfectly predict all of the situations leading to both unstressed and stressed states.

TABLE IV CONFUSION MATRIX FOR CNN RESULTS IN THE PJM INTERCONNECTION SYSTEM

| CNN | Stressed | Unstressed | Sum |
|---|---|---|---|
| Stressed | True Positive (TP) = 234 | FN (type II error) = 0 | 234 |
| Unstressed | FP (Type I error) = 0 | TN = 1687 | 1687 |
| Sum | 234 | 1687 | 1921 |

TABLE V CONFUSION MATRIX FOR ANN RESULTS IN THE PJM INTERCONNECTION SYSTEM

| ANN | Stressed | Unstressed | Sum |
|---|---|---|---|
| Stressed | TP =128 | FN (type II error) = 113 | 241 |
| Unstressed | FP (Type I error) = 106 | (TN) = 1574 | 1680 |
| Sum | 234 | 1687 | 1921 |

TABLE VI CONFUSION MATRIX FOR DECISION TREE RESULTS IN THE PJM INTERCONNECTION SYSTEM

| DT | Stressed | Unstressed | Sum |
|---|---|---|---|
| Stressed | TP = 134 | FN (type II error) = 200 | 334 |
| Unstressed | FP (Type I error) = 100 | TN = 1487 | 1587 |
| Sum | 234 | 1687 | 1921 |

## V. CONCLUSION

In this work, an online system stress assessment framework using a deep learning approach based on CNN is developed. The CNN module takes advantage of the high-speed phase angle data stream from PMUs to predict the possibility of having stress in the system, based on new post-contingency topologies. To accurately identify the severity of the stress conditions, the stress index, which is a function of power flow, was developed. The effectiveness of the proposed method in terms of accuracy and computation time has been thoroughly demonstrated using the IEEE 118-bus test system and further confirmed on the PJM interconnection system. The CNN-based module was able to automatically extract features from massive input bus phase angle data and provided both system stress status and the severity of the system stress for any OC with a highly accuracy. On top of that, the module has high accuracy to assess the security of any OC considerably faster than the traditional contingency screening techniques. The practical advantage of the developed CNN module is that it can be used as a faster alternative for RTCA in the case of failure in SE. Further, the results were compared with those from models based on ANN and DT. The CNN outperformed the other two methods in terms of accuracy. More importantly, the CNN model, in contrast to the ANN and DT models, could accurately distinguish any possible stressed states in the system.